\documentclass[10pt,conference]{IEEEtran}
\IEEEoverridecommandlockouts
\usepackage{cite}
\usepackage{amsmath,amssymb,amsfonts}
\usepackage{algorithmic}
\usepackage{graphicx}
\usepackage{textcomp}
\usepackage{xcolor}
\usepackage{framed}
\usepackage{listings}
\usepackage{float}
\usepackage{url}
\usepackage[inline]{enumitem}
\usepackage{placeins}
\usepackage{amssymb}
\usepackage{adjustbox}
\usepackage[normalem]{ulem}
\usepackage[utf8]{inputenc}

\definecolor{dkgreen}{rgb}{0,0.6,0}
\definecolor{gray}{rgb}{0.5,0.5,0.5}
\definecolor{mauve}{rgb}{0.58,0,0.82}

\begin{document}

\title{The role of \textit{slicing} in test-driven development
\thanks{This research has been sponsored by the research grants ESEIL (Finnish research agency TEKES) and TIN2014-60490-P (Spanish Ministry of Economy and Competitiveness).}
}
\author{
\IEEEauthorblockN{Oscar Dieste\IEEEauthorrefmark{1}, Ayse Tosun\IEEEauthorrefmark{2}, Sira Vegas\IEEEauthorrefmark{1}, Adrian Santos\IEEEauthorrefmark{3}, Fernando Uyaguari\IEEEauthorrefmark{4}, Jarno Kyykk{\"a}\IEEEauthorrefmark{5} and Natalia Juristo\IEEEauthorrefmark{1}}

\IEEEauthorblockA{\IEEEauthorrefmark{1}Escuela T\'ecnica Superior de Ingenieros Inform\'aticos\\
Universidad Polit\'ecnica de Madrid, Boadilla del Monte, Spain 28660\\
Email: \{odieste, svegas, natalia\}@fi.upm.es} 

\IEEEauthorblockA{\IEEEauthorrefmark{2}Faculty of Computer and Informatics\\
Istanbul Technical University, Istanbul, Turkey 34469\\
Email: tosunay@itu.edu.tr} 

\IEEEauthorblockA{\IEEEauthorrefmark{3}Faculty of Information Technology and Electrical Engineering\\ University of Oulu, Oulu, Finland 90014\\
Email: adrian.santos@oulu.fi}

\IEEEauthorblockA{\IEEEauthorrefmark{4}Empresa de Telecomunicaciones y Agua Potable de Cuenca (ETAPA), Cuenca, Ecuador 10204\\
Email: fuyaguari@etapa.net.ec}

\IEEEauthorblockA{\IEEEauthorrefmark{5}Ericsson, Jorvas, Finland 02420\\
Email: jarno.kyykka@ericsson.com}}

\maketitle

\begin{abstract}
Test-driven development (TDD) is a widely used agile practice. However, very little is known with certainty about TDD's underlying foundations, i.e., \textit{the way TDD works}. In this paper, we propose a theoretical framework for TDD, with the following characteristics: 
\begin{enumerate*}[label={\alph*)}]
\item Each TDD cycle represents a vertical slice of a (probably also small) user story,
\item vertical slices are captured using contracts, implicit in the developers' minds, and
\item the code created during a TDD cycle is a sliced-based specification of a code oracle, using the contracts as slicing pre/post-conditions
\end{enumerate*}. We have checked the connections among TDD, contracts and slices by means of a controlled experiment conducted in industry.
\end{abstract}


\section{Introduction}

Test-driven development (TDD) is one of the best known Extreme Programming practices \cite{beck2004extreme}, later adopted by Agile methodologies. In the last State of Agile survey \cite{collabnet2018agile}, 35\% of agile practitioners reported that they use TDD regularly (although the percentage is declining over time; in previous surveys, TDD adoption was around 50\% \cite{collabnet2009agile}). 

Not surprisingly, TDD received substantial attention in Software Engineering (SE) research (e.g., see Rafique and Mi{\v{s}}i\'c's \cite{rafique2013effects} literature review) and practice (countless of web pages are devoted to TDD). However, in contrast to the apparent interest in TDD, very little is known with certainty about \textit{how TDD works}. There are vivid discussions about the role of the tests in TDD, the design/quality assurance character of TDD, or even whether TDD improves quality and productivity \cite{karac2018what}.

This research \textbf{aims to clarify the underlying foundations of TDD}. In our opinion, TDD is a correctness technique, based on contracts. Contracts are operationalized by means of  test cases. TDD programming is a contract-development process based on vertical slices of a problem specification.

This paper makes \textbf{two contributions}:
\begin{enumerate}
\item We propose a theoretical framework connecting the concepts of TDD (in particular, unit tests and TDD cycles), contracts, and slicing (in two flavors: vertical slices, and specification-based slices).
\item We provide evidence about the soundness of the theoretical framework by means of an industry experiment.
\end{enumerate}

The paper is organized as follows: Section~\ref{sec:background} reviews the key concepts used in this research. Section~\ref{sec:theory} reports the theoretical framework. Section~\ref{sec:worked} provides a relatively long worked example using the ideas introduced in Section~\ref{sec:theory}. Section~\ref{sec:experiments} validates the theoretical framework empirically. Finally, Section~\ref{sec:worked} highlights the key results and suggests some future lines of research.

\section{Background}\label{sec:background}

\subsection{User stories}

\textit{User stories} are the one of the primary practices in Extreme Programming \cite{beck2004extreme}. They are ''units of customer-visible functionality'' that enable early estimation and provide business benefit. User stories are often too ''big'' or ''large'', i.e., they imply a substantial amount of work to be done. Large stories are difficult to estimate and, if they are sufficiently big, cannot be accommodated in one sprint. In either case, user stories should be split \cite{cohn2005agile}. 

User stories conform the INVEST\footnote{Independent, Negotiable, Valuable, Estimable, Small and Testable.} model \cite{wake2003INVEST}. The \textit{valuable} dimension restricts the possible splitting strategies, e.g., \textit{functional decomposition} is inadequate because the resulting modules are unlikely to provide business value by themselves \cite{leffingwell2010agile,wake2003INVEST}. \textit{Vertical slicing} is the almost universally recommended approach \cite{cohn2005agile,leffingwell2010agile,millett2011pro} to split user stories (with few exceptions, e.g., \cite{kerievsky2011return}). 

\subsection{Vertical slicing}\label{sec:background:vertical-slicing}

Vertical slicing is frequently described as to ''slice'' vertically (thus the name) through the layers of a ''cake'' \cite{wake2003INVEST}. When user stories are vertically sliced, a functionally coherent and demostrable (sub-)product is developed from user interface to database \cite{ratner2011vertical}, crossing all architectural layers. 

User stories are sliced using heuristics referred to as \textit{patterns}. The most popular patterns have been compiled/proposed by Richard Lawrence \cite{lawrence2012new}, which have found their way into the technical literature \cite{leffingwell2010agile}. These patterns operate within and along business process boundaries (e.g., the \textit{operations} and \textit{business rules variations} patterns), or project management considerations (e.g., \textit{simple/complex}, \textit{major effort}). Patterns are not related to technical issues, because user stories are expressions of user needs, i.e., requirements. The patterns with a stronger technical flavor are the \textit{interface variations} and \textit{defer performance} patterns, but in both cases they make reference to external observable behavior. 

The ''cake'' metaphor suggests that the concept of vertical slice lacks formal definition. It is rather easy to find references to ''vertical slicing'' in internet blogs\footnote{For instance, some popular blogs are:
\begin{itemize}
\item \url{http://blogs.adobe.com/agile/2013/09/27/splitting-stories-into-small-vertical-slices/}
\item \url{https://agileforall.com/vertical-slices-and-scale/}
\item \url{https://www.deltamatrix.com/horizontal-and-vertical-user-stories-slicing-the-cake/}
\end{itemize}} and technical books \cite{cohn2005agile,leffingwell2010agile,millett2011pro}, i.e., practitioner-oriented literature. In turn, the idea of ''vertical slice'' has not progressed into the academic literature\footnote{E.g., the reference \cite{ratner2011vertical} above is one of the few examples of indexed publications addressing vertical slicing.}. It is thus not surprising that the properties of vertical slices are not well defined. One exception is that vertical slices are expected to comply with the INVEST model \cite{ansari2016splitting}, i.e., they are small-scale user stories independently \textit{testable} \cite{ekas2013being,virgo2016understanding}. The associated tests are non-brittle, i.e., they are fairly stable, regardless of development considerations, as long as the user needs keep stable as well.

\subsection{Test-driven development}\label{sec:background:tdd}

Test-driven development (TDD) is an evolution of the ''test-first'' practice in Extreme Programming \cite{beck2004extreme}. According to its creator, K. Beck, TDD is not a testing but a logical design technique \cite{beck2001aim}. In TDD, code is written following a ''rhythm'' \cite[pp. 1]{beck2003test}, i.e., a cyclical development process:
\begin{enumerate}
\item Quickly add a test.
\item Run all tests and see the new one fail.
\item Make a little change.
\item Run all tests and see them all succeed.
\item Refactor to remove duplication.
\end{enumerate}

Beck argues that TDD is \textit{a way of managing fear during programming}. When engaged in complex tasks, programmers become tentative \cite[pp. xi]{beck2003test}; they perform too much up-front design, leading to higher complexity and slower progress. In turn, tests make programmers to do something concrete and quickly.

K. Beck does not give clear directions regarding the tests to write. He says \textit{''write a list of all the tests you know you will have to write[...]. First, put on the list examples of every operation that you know you need to implement. Next, for those operations that don't already exist, put the null version of that operation on the list''} \cite[pp. 126]{beck2003test}. Apparently, K. Beck means the vertical slices described in the previous Section.

Other authors emphasize the testing character of TDD. For instance, T. Dings\o yr et al. \cite[pp. 196]{dingsyr2010agile} describe TDD as \textit{''a safety net of test covering the entire application. This safety net means that developers can confidently change code in response to changing requirements since regression errors will be caught immediately by existing tests''}. Some authors as H. Wayne \cite{wayne2018tdd} go beyond this characterization, and consider that TDD is a \textit{correctness} technique, in line with (although different than) \textit{design by contract}\cite{meyer1992applying}, among others.

\subsection{Slicing}\label{sec:background:slicing}

\textit{Slicing} is a program decomposition technique proposed by  M. Weiser \cite{weiser1981program}. A program \textit{slice} is a subset (of the lines of code) of a program, which preserves a ''projection'' of its ''behavior''. The \textit{behavior} of a program is defined as a list of pairs $(n_i,s_i)$, where $n_i$ is a statement (/line of code) and $s_i$ the set of variables modified by that statement (\textit{state trajectory}). In other words, the behavior of a program is a trace of the program execution. A \textit{projection} of the behavior is a subset of the state trajectory (i.e., a trace of some subset of variables). 

Projections are defined using slicing criteria. Typically, slicing criteria are pairs $<line\ number, variables>$. The slicing criteria ''throw out'' those statements that do not affect, or are not affected, by the slicing criteria \textit{variables} at the \textit{line number}. The remaining statements (which do have an influence on the slicing criteria variables) make up the program slice. 

\subsection{Specification-based slicing}

There are many slicing approaches (see, e.g., \cite{tip1995survey,delucia2001program}), and not all of them rely on a pairs $<line\ number, variables>$ as slicing criteria. In particular, programs can be sliced with respect to a precondition $\{P\}$ and/or a postcondition $\{Q\}$ \cite{chung2001program,comuzzi1996program}, i.e., a program specification. 

In terms of Hoare's triples \cite{hoare1969axiomatic,gries1981science}, a specification-based slice of a program $S$ is another program $S\prime$ such that $\{P\}S\{Q\} \implies \{P\}S\prime\{Q\}$ and $S\prime$ can be obtained removing statements from $S$. A specification-based slice is a regular slice, although the preserved behavior can be easily understood in terms of a simple (without invariants or \textit{rescue} procedures) contract $\{P\}\{Q\}$ \cite{meyer1992applying}.

\section{Theoretical framework}\label{sec:theory}

The \textit{nature} of TDD, i.e., its essential qualities, have been discussed at length in blogs and discussion boards with little agreement. Our position is that TDD is a correctness technique, much in line with H. Wayne \cite{wayne2018tdd}:
\begin{enumerate*}[label={\alph*)}]
\item unit tests are mechanisms that provide confidence on the code complies with a given contract, and
\item contracts develop throughout the TDD cycles
\end{enumerate*}. The code is up to the developer, i.e., TDD is not a programming or design technique, although TDD can suggest programming and design decisions\cite{beck2001aim}. In the following Sections we formalize these ideas, which represent the \textbf{first contribution} of this paper.

\subsection{A test case is a concrete instance of a contract}\label{sec:theory:contract}

A test case defines an input-output relationship. However, developers do not write test cases having a concrete relationship in mind \textit{only}. In turn, the test case represents a concrete instance of the execution of a (not yet written) algorithm, subjected to some restrictions (pre/post-conditions) $\{P\}C\{Q\}$. In other words, a test case is a concrete instance of a simplified contract (without invariants or \textit{rescue} procedures).

$P$ and $Q$ are not explicit, but implicit in the developer's mind. For instance, when a developer writes the test case\footnote{The examples will be written in C++. The unit testing framework is \textit{Boost Test} \cite{rozental2018test}. The choice of language/framework was made by the company in which the experiments were conducted (see Section~\ref{sec:experiments}).}:
\begin{lstlisting}
BOOST_AUTO_TEST_CASE (return_a_when_a_greater_b)
{
	const int a = 2;
	const int b = 1;
	
	BOOST_CHECK_EQUAL(a, max(a, b));
}
\end{lstlisting}
for the \lstinline{max()} function\footnote{The example is taken from Gries \cite[pp. 108]{gries1981science}.}, she is not thinking specifically in the values $a=2$ and $b=1$, but in the following contract:

\begin{center}
$P:\{a > b\}$
\begin{lstlisting}
int max(int a, int b){
	if (a > b) return a;
}
\end{lstlisting}
$Q:\{a > b\ \land\ $\lstinline{max}$= a\}$
\end{center}

\subsection{Regression tests provide an informal proof of correctness}

A single test case cannot verify that the code works for the the entire set of states represented by the contract. The regression\footnote{TDD does not have a precise terminology. We designate these tests as 'regression tests' because they do not fail when the code is correct.} test(s) explore(s) different regions of the set of states, increasing the developer's confidence on the code, i.e., they play the same role than a proof of correctness. For instance, the developer can write the following test:
\begin{lstlisting}
BOOST_AUTO_TEST_CASE (return_a_when_a_greater_b_regression)
{
	const int a = -1;
	const int b = -2;
	
	BOOST_CHECK_EQUAL(a, max(a, b));
}
\end{lstlisting}
to make sure that the \lstinline{max()} function works for negative numbers.

These regression tests may, or may not, have an influence on the contract. It is hard to find out, due to its implicit character. If the pre-condition in Section~\ref{sec:theory:contract} above were as stated, the new test case would not imply any change to the contract. However, it is possible that the developer was thinking in terms of positive numbers, i.e. the contract were:
\begin{center}
$P:\{a > b \land a > 0 \land b > 0\}$\\
\lstinline{max(a, b)}\\
$Q:\{a > b\ \land\ $\lstinline{max}$= a\}$
\end{center}

In such a case, the contract would be modified by the new test case. Common sense suggests that the changes aim to \textit{weaken} the the pre-condition, i.e., ensure that the program works correctly for a wider input set. This is an instance of a more general evolutionary process, described in the next Section.

\subsection{New tests cases modify the existing contract}\label{sec:theory:new}

When a developer adds a new test case (not a regression test), she has the intention of adding new behavior to the existing code. For instance, the following test:
\begin{lstlisting}
BOOST_AUTO_TEST_CASE (return_b_when_a_less_or_equal_b)
{
	const int a = 3;
	const int b = 4;
	
	BOOST_CHECK_EQUAL(b, max(a, b));
}
\end{lstlisting}
is a concrete instance of the contract:
\begin{center}
$P:\{a \leq b\}$
\lstinline{max(a, b)}
$Q:\{a \leq b\ \land\ $\lstinline{max}$= b\}$
\end{center}
as we have already described in Section~\ref{sec:theory:contract}. If this new contract were implemented independently of the first one, the situation would be similar to what is displayed in Table~\ref{tab:theory:interactions}.

\begin{table}[h!]
\caption{Two TDD cycles, and the corresponding contracts}
\label{tab:theory:iterations}
\begin{tabular}{c|c}
First TDD cycle & Second TDD cycle \\ \hline

$P:\{a > b\}$ & $P\prime:\{a \leq b\}$\\ % end of 1st row

\begin{lstlisting}
int max(int a, int b){
	if (a > b) return a;
}
\end{lstlisting}

&

\begin{lstlisting}
int max(int a, int b){
	if (a <= b) return b;
}
\end{lstlisting}\\ % end of 2nd row

$Q:\{a > b\ \land\ $ \textit{max} $=a\}$ & $Q\prime:\{a \leq b\ \land\ $ \textit{max} $=b\}$ % end of 3rd row

\end{tabular}
\end{table}

Of course, code is not developed this way. Subsequent TDD cycles build upon previous ones, making the code increasingly more abstract. In practice, the outcome of a second TDD cycle shall be:
\begin{center}
$P\prime\prime:\{(a > b) \lor (a \leq b)\} \equiv \{\top\}$\\
\begin{lstlisting}
int max(int a, int b){
	if (a > b) return a;
	else return b;
}
\end{lstlisting}
$Q\prime\prime:\{(a > b\ \land $\ \lstinline{max}$=a) \lor (a \leq b\ \land$\ \lstinline{max}$=b)\}$
\end{center}

The new contract $\{P\prime\prime\}\{Q\prime\prime\}$ is the \textit{union} (logical-OR) of the contracts $\{P\}\{Q\}$ and $\{P\prime\}\{Q\prime\}$. The effect of this union is that both the pre/post-conditions are \textit{weakened}, i.e., the code is able to handle a larger number of states (and the corresponding inputs and outputs), in line with the constructive nature of the TDD cycles.
%We have observed that this is not always the case; in many cases, the postconditions do not change, i.e., \textit{only the preconditions are weakened} (see next Section).

In practice, TDD programmers can use different mechanisms to develop the contracts throughout the TDD cycles. For instance, contracts can be refined, i.e., \textit{strengthened}. However, the \textit{union} guarantee that previous tests are not affected by later TDD cycles, i.e., the test cases corresponding to the $\{P\}\{Q\}$ and $\{P\prime\}\{Q\prime\}$ contracts are also concrete instances of the $\{P\prime\prime\}\{Q\prime\prime\}$ contract. 

\subsection{The code generated during a TDD cycle is a contract-based slice}

Table~\ref{tab:theory:iterations} clearly shows that the code written for each contract is a ''projection'' of the \textit{complete} \lstinline{max()} code. More specifically, such projection is a specification-based slice \cite{chung2001program,comuzzi1996program}. The contract $\{P\}\{Q\}$ gives the code in the left-hand side of Table~\ref{tab:theory:iterations}, by means of precondition-based slicing as shown in Listing~\ref{listing:precondition-based-slice}. The right-hand side of Table~\ref{tab:theory:iterations} can be likewise obtained from the contract $\{P\prime\}\{Q\prime\}$.

\begin{center}
\begin{lstlisting}[mathescape=true,caption={Precondition-based slice},label={listing:precondition-based-slice}]
$P:\{a > b\}$
int max(int a, int b) {
	if(a > b) 
		return a; // max := a
		$\{a > b\ \land\ $max$=a\}$
	¡else¡
		¡return b;¡
		$\{a > b\}$ // The precondition does not change, so the else branch can be removed
}
$Q:\{a > b\ \land$ max $=a\}$
\end{lstlisting}
\end{center}

We do not claim that TDD programmers create precise slices out of a \textit{code oracle}. Programming is a creative activity, and only in very simple cases, e.g., the \lstinline{max()} function, or the \lstinline{div()} function worked out in Section~\ref{sec:worked}, such 1:1 relationship is possible. In more complex cases, we can only expect a semantic relationship, i.e., given the contract $\{P\}\{Q\}$, the \textit{code oracle} $C$ and the programmers code $S$:

\begin{center}
$\{P\}S\{Q\} \implies \{P\}C\{Q\}$
\end{center}

\subsection{A contract-based slice is a (nano) vertical slice}\label{sec:theory:slice}

A contract-based slice has the same properties than a vertical slice:
\begin{itemize}
\item A slice represents a \textit{functionally coherent and demostrable (sub-)product}, e.g., the code in Table~\ref{tab:theory:iterations} is fully functional, and each TDD cycle provides a partial solution to the customer need\footnote{We ask for the readers' indulgence.}.
\item A slice complies with the INVEST model, in particular the \textit{testable} property. Tests become robust due to the OR-composition of contract-based slices. Later TDD cycles do not break previous tests, i.e., $\{P\}\{Q\} \subset \{P\prime\}\{Q\prime\} \subset \{P\prime\prime\}\{Q\prime\prime\} \subset \dots$.
\end{itemize}

\section{Worked example}\label{sec:worked}

In this section, we ground the theoretical concepts introduced in the previous Section. We  code %three 
a simple function using TDD, and we show that the most plausible solution is driven by contract-based slices. 

%The first example is the \lstinline{swap()} function\footnote{The example is taken from Gries \cite[pp. 102]{gries1981science}.}. \lstinline{swap()} is probably one of the simpler conceiving examples and can be used to illustrate the key concepts, e.g., iterative post-condition, without unnecessary complexity. 

%The second example is the \lstinline{div()} function\footnote{The example is taken from from Gries \cite[pp. 1]{gries1981science}.}. This function is more complex than \lstinline{swap()}, e.g., it includes a loop invariant, and the post-condition can be asserted since the first TDD cycles.

%Finally, the third example is Robert Martin's (Uncle Bob) \lstinline{prime_factors()} kata \cite{martin2005prime}. This example is somewhat reiterative (there are no substantial differences between \lstinline{div()} and \lstinline{prime_factors()}); however, this kata is interesting because it comes from a reliable, independent source, yet Robert Martin's solution can be also read in terms of PBSs.

%\input{swap-example}

%\subsection{The \lstinline{div()} function}\label{sec:worked:div}

The proposed example is the \lstinline{div()} function. \lstinline{div()} calculates the integer division $x \over y$, giving the quotient in $q$ and the remainder in $r$\footnote{The example is taken from from Gries \cite[pp. 1]{gries1981science}. Notice that $x \geq 0$ and $y > 0$.}. In Tables~\ref{tab:div-listing1} and \ref{tab:div-listing2} (located at the end of the paper), we show how this code can be obtained using TDD, and how each TDD cycle matches the concept of contract-based slice\footnote{We use Robert Martin's layout (tests at the left, \lstinline{div()} code at the right) to display the progress of the programming session. When appropriate, we show in the center the contract-based slice, following Chung et al. \cite{chung2001program}, using Gries' notation \cite{gries1981science}.}. 

The \textit{oracle code} appears in the last row of Table~\ref{tab:div-listing2}. The programming session has 9 TDD cycles (one per row). Let's start with Table~\ref{tab:div-listing1}:

\begin{enumerate}
\item In the first TDD cycle, a simple case ($2 \over 2$) is proposed. The code behavior is faked using constants, according to the ''Fake it'' strategy \cite[pp. 151]{beck2003test}. The contract-based slice is computed using an ''unrolled'' version of the \textit{div} algorithm, i.e., the loop is unrolled as a sequence of \lstinline{if}s. Notice that the ''rolled'' version of \lstinline{div()} becomes clear only later in the programming session.

The slice is computed using the precondition $P:\{x = 2 \land y = 2\}$, for two reasons:
\begin{enumerate*}[label={\alph*)}]
\item the concrete character of the precondition (variables are assigned constant values) makes the computation trivial, and
\item the precondition-based slice seems to be aligned with the way that the code is created using TDD (the values of the variables \lstinline{x} and \lstinline{y} suggest the code to write)
\end{enumerate*}.

The slice does not look like the code on the right side of the table\footnote{In Section~\ref{sec:theory}, we already commented that a 1:1 relationship between the slice code and the TDD code cannot be expected, due to the creative character of programming.}. The differences are probably due to some ''shortcuts'' present in the developer's mind. Notice that the code snippet:

\begin{lstlisting}[mathescape=true]
$\{x = 2 \land y = 2\ \land t = 2 \land q = 0\}$
	if(t >= y) { // unrolled loop.
		t = t - y;
		(*q)++;
	}
$\{x = 2 \land y = 2\ \land t = 0 \land q = 1\}$
\end{lstlisting}

can be simplified thanks to the concrete character of the preconditions:
\begin{enumerate*}[label={\alph*)}]
\item the \lstinline{if()} can be removed, and
\item the variables replaced with constants.
\end{enumerate*}. This gives the code:

\begin{lstlisting}[mathescape=true]
$\{x = 2 \land y = 2\ \land t = 4 \land q = 0\}$
	*q = 1;
$\{x = 2 \land y = 2\ \land t = 2 \land q = 1\}$

\end{lstlisting}

which is precisely the outcome of the 1\textsuperscript{st} TDD cycle.

\item The code is generalized to handle the $x = 2 \land x = 4$ cases.

\item The developer wishes to generalize the algorithm to handle arbitrary even numbers, i.e.,  implement the contract $P:\{x = 2^n \land n > 0 \land y = 2\}$. However, the code structure is not yet clear, so that a new baby step is given before abstracting the code; the precondition is restricted to the $x = 2 \land x = 4 \land x = 6$ cases.

\item The repetitive structure of the code in the previous step suggests that the sequence of \lstinline{if}s can be replaced by a loop. The regression tests succeed. The developer is confident that the code works for $x = 2^n \land n > 0$. A triangulation test \cite[pp. 153]{beck2003test} would increase the confidence even further, but it looked unnecessary.

There are some differences between the slice and TDD code. Again, such differences can be reconciled attending to the concrete character of the pre/post conditions. For instance, 
given the postcondition $Q$ in the code snippet below:

\begin{lstlisting}[mathescape=true]
	*r = t;
}
$Q:\{0 \leq r < y \land x=y*q+r \land r = 0\}$
\end{lstlisting}

the developer can simplify the assignment \lstinline{*r = t;} as \lstinline{*r = 0;}. The assignment \lstinline{*q = 1;} can be also obtained from the slice, as indicated above in the TDD cycle 1.

It is also noticeable that the contract-based slices computed so far were precondition-based slices. From this TDD cycle on, the precondition ($P:\{x = 2^n \land n > 0 \land y = 2\}$) becomes too abstract as to suggest which code to write. In turn, the postcondition triggers a reasoning process that leads to performing small changes in the code, in the direction of the final solution. In parallel, the postcondition-based slices get easier to compute (than the precondition-based slices). This observation is collateral, but it suggests characteristics of the contract-based slicing worth exploring in future research.

\end{enumerate}

The most relevant steps towards the implementation of the \lstinline{div()} function have been already taken. The remaining TDD cycles in Table~\ref{tab:div-listing2} simply round up the work, and have less value from the viewpoint of examining the relationship between contract-based slices and TDD-generated code:

\begin{enumerate}

\setcounter{enumi}{4}

\item The developer wonders if the code is able to handle the case $x = 2^0$. That implies that the statement \lstinline{*q = 1;} is rewritten as \lstinline{*q = 0;}.

\item The developer realizes that the code is able handle arbitrary exact integer divisions (precondition $P:\{(x = y * q\}$). 

\item A triangulation test double checks that the code performs exact integer divisions correctly.

\item The developer writes a test for a non-exact integer division, i.e., the precondition $P:\{x=y*q+r\}$. Only a change to the condition of the \lstinline{while} loop is necessary to pass the test.

\item Apparently, the code for \lstinline{div()} is completed. A triangulation test, exploring an edge case, confirms that the code is working properly.

\end{enumerate}

\section{Experimental evaluation}\label{sec:experiments}

The \textbf{second contribution} of this paper is a controlled experiment aimed at testing the role of slicing in TDD. In Section~\ref{sec:theory}, we have postulated a theoretical framework connecting the concepts \textit{TDD-contracts-slices}. Several dimensions of this framework can be explored; in this paper, we focus in the foundational one, i.e., whether we can justify, on an empirical basis, a connection between TDD programming and contract-based slices. Connections between slicing and programming/maintenance have been established already, e.g., \cite{weiser1982programmers,binkley2001implementation,sinha1999system}. The research question can be stated as:  

\begin{framed}
RQ. Does slicing have an influence on TDD?
\end{framed}

By ''influence'', we mean an effect on any relevant construct. The code \textit{External quality} has been frequently used in TDD research (e.g., see Rafique and Mi{\v{s}}i\'c's \cite{rafique2013effects} literature review). Accordingly, we investigated whether slicing improves (or worsens) \textit{External quality} when using TDD. The experiment will be described using Jedlitschka et al.'s reporting guidelines \cite{jedlitschka2008reporting} as length permits.

\subsection{Experimental planning}

\subsubsection{Summary}\label{sec:summary}

The experiment was conducted in Ericsson Finland. The experiment was part of a 3-day TDD training seminar, delivered by instructors from the universities of Oulu and Polit\'ecnica de Madrid.

The experiment has been conducted in two phases. In the 1\textsuperscript{st} phase, a set of programmers performed a programming task using interactive test-last development (ITLD) \cite{madeyski2005preliminary}. ITLD involves writing production and testing code in parallel; however, tests are not created before production code as in TDD. The ITLD strategy is in widespread use in industry \cite{williams2009effectiveness}. 

In the 2\textsuperscript{nd} phase, the same programmers performed a different task using TDD. They had not previous professional experience in TDD. In between phases 1 and 2, the programmers went through specific TDD training.

The company decided the programming language and environment: The code was written in C++ using \textit{Boost Test} and Eclipse CDT. The quality of the generated code was measured using test suites. Finally, we checked whether slicing had an effect on the \textit{External quality} of the code generated using the ITLD/TDD programming strategies.

\subsubsection{Independent variables} 

There are two factors: 
\begin{enumerate*}[label={\alph*)}]
\item \textit{Programming strategy}, and 
\item \textit{Slicing}
\end{enumerate*}. 
Each factor has two levels. The \textit{Programming strategy} factor has the ITLD and TDD levels.  The \textit{Slicing} factor has the levels \textit{yes} and \textit{no}. \textbf{We explain in Section~\ref{sec:experiments:materials} below how the \textit{yes/no} levels have been operationalized}.

\subsubsection{Response variables} We used two test suites, one for each exercised task. The \textit{External quality} was calculated on the basis of the passed/failed assertions as:

\begin{equation}\label{eq:qlty}
QLTY = \frac{\#Assert(Pass)}{\#Assert(All)} 
\end{equation}

QLTY is a percentage. The possible scores range from 0 to 100. The test suites are available at \url{https://goo.gl/Pprgdm}. \textbf{In case of acceptance, the files will be placed in a permanent URL}.

\subsubsection{Hypothesis} We have posed 2 hypotheses, each one corresponding to a phase of the experiment. For the 2\textsuperscript{nd} phase (TDD), the hypothesis is:

\setlist[enumerate,1]{start=0}
\begin{itemize}[label={}]
\item 
	\begin{enumerate}[label*=H1$_\arabic*$]
  		\item Slicing does not have an effect when using TDD: \\
  	   		  $QLTY_{Slicing = no} = QLTY_{Slicing = yes}$
  		\item Slicing has an effect when using TDD:\\ 
			  $QLTY_{Slicing = no} \neq QLTY_{Slicing = yes}$
	\end{enumerate}
\end{itemize}

This hypothesis targets whether slicing has an effect on TDD. If the null hypothesis is rejected, that would imply that slicing makes an effect on TDD's \textit{External quality}. We expect that such effect will be positive, although we cannot rule out that slicing is detrimental to TDD; thus, the  alternate hypothesis is 2-tailed.

For the 1\textsuperscript{st} phase (ILTD programming), we pose exactly the same hypothesis: 

\begin{itemize}[label={}]
\item 
	\begin{enumerate}[label*=H2$_\arabic*$]
  		\item Slicing does not have an effect when using ITLD: \\
  	   		  $QLTY_{Slicing = no} = QLTY_{Slicing = yes}$
  		\item Slicing has an effect when using ILTD:\\ 
			  $QLTY_{Slicing = no} \neq QLTY_{Slicing = yes}$
	\end{enumerate}
\end{itemize}
\setlist[enumerate,1]{start=1}

Hypothesis H2 functions as a ''control'' to H1, i.e., it is possible that slicing would not be specifically linked to TDD, and have some effect in other programming strategies such as ITLD. In fact, there is available evidence that contracts increase code \textit{External quality} \cite{muller2002two}. If H2 were rejected, then slicing\footnote{Notice that the the theoretical framework in Section~\ref{sec:theory} states that TDD cycles are contract-based slices, but no claims are made about contracts in general.} would represent a general conceptual approach to programming, but not an underlying foundation of TDD.

\subsubsection{experimental subjects} 20 professional programmers. Table~\ref{tab:experiments:experience} summarizes the programmers' experience. All of them are inexperienced in unit testing and TDD.

\begin{table}[]
\caption{Programmers' experience}
\centering
\begin{tabular}{ll}
\hline
Experience                            & \# \\ \hline
No experience (\textless{}2 years)    & 2  \\
Novice (2-\textless{}=5 years)        & 7  \\
Intermediate (5-\textless{}=10 years) & 10 \\
Expert (\textgreater{}10 years)       & 2  \\ \hline
\end{tabular}
\label{tab:experiments:experience}
\end{table}

\subsubsection{Tasks} We used two experimental tasks, for two reasons:
\begin{enumerate*}[label={\alph*)}]
\item The same subjects participate in the 1\textsuperscript{st} and 2\textsuperscript{nd} experimental phases.
\item \textbf{Tasks often interact with the experimental factors}. Using more than one tasks increases the confidence in the experimental results.
\end{enumerate*}

The tasks are Bowling ScoreKeeper (BSK) and MarsRover API (MR). The goal of BSK is to calculate the score of a single bowling game. BSK is a modified version of Robert Martin's Bowling Scorekeeper \cite{martin2005bowling}. This task is popular in the agile community, e.g., \cite{codingdojo}, and was used in previous TDD experiments, e.g. \cite{erdogmus2005effectiveness,fucci2013replicated,williams2003test}. The goal of MR is controlling the movement of a fictitious vehicle on a grid with obstacles. MR is also a popular in the agile community, e.g., \cite{kata-log}.

\subsubsection{Experimental materials}\label{sec:experiments:materials} The tasks BSK and MR were provided to the experimental subjects in two different layouts:
\begin{enumerate*}[label={\alph*)}]
\item a natural language, 1/2-page specification, including examples, and
\item a sliced specification
\end{enumerate*}. 

The sliced/non-sliced specifications \textbf{represent the operationalization of the \textit{Slicing} factor}. As indicated above in Section~\ref{sec:summary}, the programmers participating in the experiment do not have TDD experience. It is highly unlikely that the short training course give the programmers the necessary skills for successful vertical slicing. In our opinion, the programmers that receive a non-sliced specification \textit{probably cannot figure out the right slices by themselves}.

In turn, the sliced specification contains the right vertical slices. The programmers that receive a sliced specification are required to use the slices during the coding session. If \textit{Slicing} had an influence on ITLD/TDD, it would show up in the code generated using a sliced specification.

A sliced specification is composed by a series of steps. Each step corresponds to one TDD cycle. For each step, we provide an explanation of the behavior to code (a textual version of the contract), and one example that can be used as test case. A non-sliced specification is simply a textual description of the function to code.

All specifications (sliced and non-sliced) are available at \url{https://goo.gl/MFUQpk}. \textbf{In case of acceptance, the files will be placed in a permanent URL}.

\subsubsection{Design} Both the 1\textsuperscript{st} and 2\textsuperscript{nd} phases were designed as 2\textsuperscript{2} factorial designs. Experimental subjects were randomly assigned. Tasks were switched between the 1\textsuperscript{st} and 2\textsuperscript{nd} phases. Specifications were counterbalanced in the 2\textsuperscript{nd} phase.

\subsubsection{Analysis procedure} The analysis has been conducted using a 2-way ANOVA. We have used R \cite{R}, and the packages \textit{car} \cite{car}, \textit{effsize} \cite{effsize}, \textit{fsa} \cite{fsa}, \textit{lmerTest} \cite{lmertest}, \textit{xtable} \cite{xtable} and \textit{xlsx} \cite{xlsx}.

\subsection{Experimental execution}

The 3-day seminar ran from 08:30 to 16:30 approximately. During the 1\textsuperscript{st} day of the seminar, the subjects received training in unit testing and \textit{Boost Test}. At the beginning of the 2\textsuperscript{nd} day, the ITLD programming session was conducted. 

The 2\textsuperscript{nd} and 3\textsuperscript{rd} days, the subjects received training in TDD, including several worked examples. The TDD programming session was conducted at the end of the 3\textsuperscript{rd} day.

The sessions were 2 hours long. The subjects could make any questions regarding the tasks or programming environment. None subject complained about time pressure.

21 subjects signed up, but one of them didn't show up. There were no drop-outs during the course, neither substantial delays, e.g., due to late arriving subjects. The exceptions were: 
\begin{enumerate*}[label={\alph*)}]
\item Due to a schedule misunderstanding, the seminar started $1 \over 2$ late the 1\textsuperscript{st} day, and
\item in one case one subject left earlier due to a child sick leave
\end{enumerate*}.

\FloatBarrier

\subsection{Analysis}

The raw data and the R code are available at \url{https://goo.gl/hFuzPx}.  \textbf{In case of acceptance, the files will be placed in a permanent URL}.

\subsubsection{ANOVA assumptions} The Levene test confirms that the data holds the heteroskedasticity assumption for both the ITLD ($F=0.92,\ df=3,\ p-value=0.45$) and TDD sessions ($F=1.02,\ df=3,\ p-value=0.41$). In the case of normality, the Shapiro-Wilk test is boderline for both the ITLD ($W=0.91,\ p-value=0.06$) and TDD ($W=0.9,\ p-value=0.04$) sessions. Notice that the ANOVA is robust to non-normality \cite{glass1972consequences, lix1996revisited}, but we will discuss this threat to validity in Section~\ref{sec:experiments:threats}.

\subsubsection{ITLD phase} Table~\ref{tab:experiment:itld:descriptives} reports the descriptive statistics for the ITLD session. QLTY scores are much the same for every combination of levels ($min: 20.26,\ max: 42.53$). Standard deviations are roughly proportional to the means. None factor level stands out as particularly influential on QLTY. 

% latex table generated in R 3.3.2 by xtable 1.8-2 package
% Thu Aug 23 16:00:49 2018
\begin{table}[ht]
\centering
\caption{Descriptive statistics for the ITLD session} 
\label{tab:experiment:itld:descriptives}
\begin{tabular}{rllrrr}
  \hline
 & SLICING & TASK & n & mean & sd \\ 
  \hline
1 & no & bsk & 6.00 & 20.26 & 25.44 \\ 
  2 & yes & bsk & 4.00 & 25.77 & 37.00 \\ 
  3 & no & mr & 5.00 & 42.53 & 45.13 \\ 
  4 & yes & mr & 5.00 & 33.79 & 21.94 \\ 
   \hline
\end{tabular}
\end{table}

Table~\ref{tab:experiment:itld:analysis} shows the results of the statistical analysis. H1\textsubscript{0} cannot be rejected. The programmers who used a sliced specification did not generate higher quality code (Cohen's $d = 0.005$, confidence interval $(-0.94, 0.95)$). 

\begin{figure}
\centering
\includegraphics[width=1.7in]{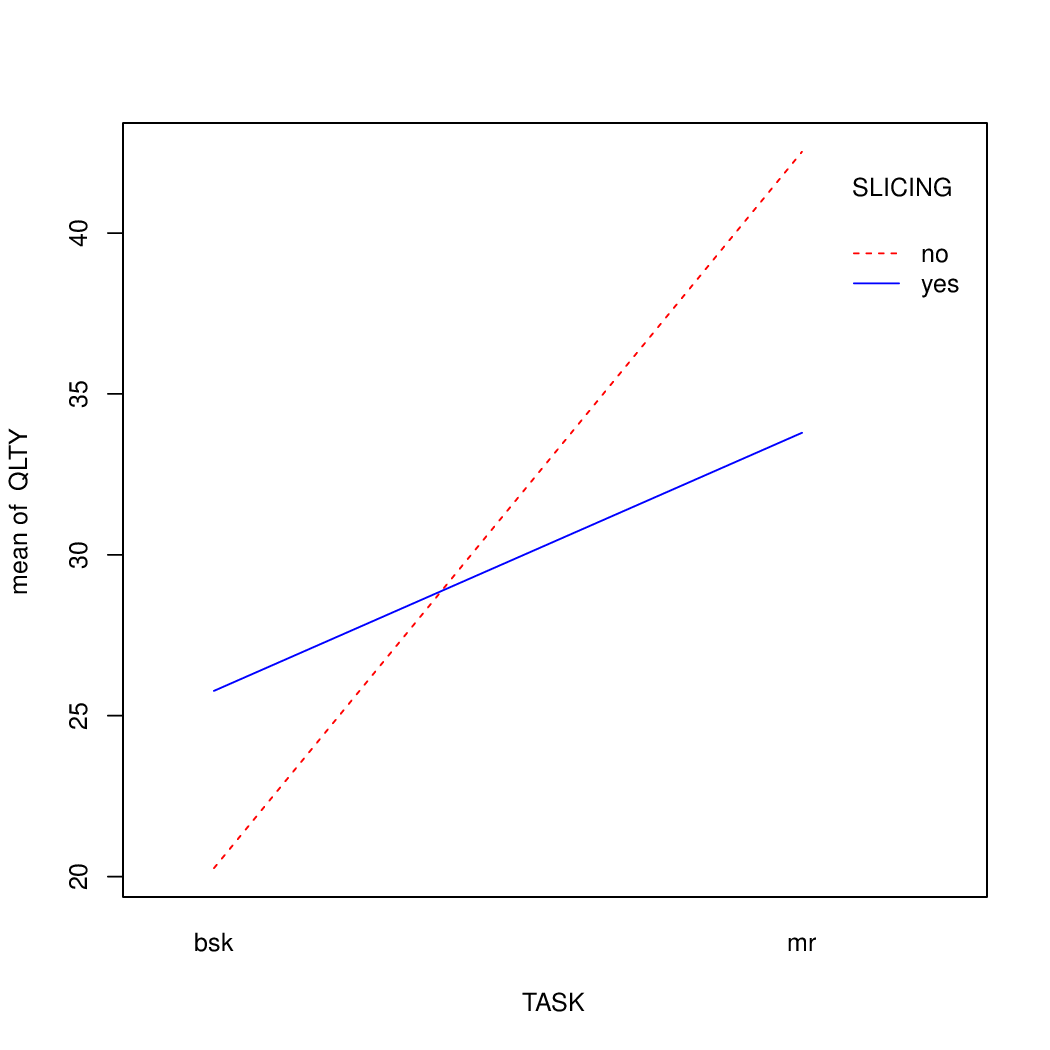}
\includegraphics[width=1.7in]{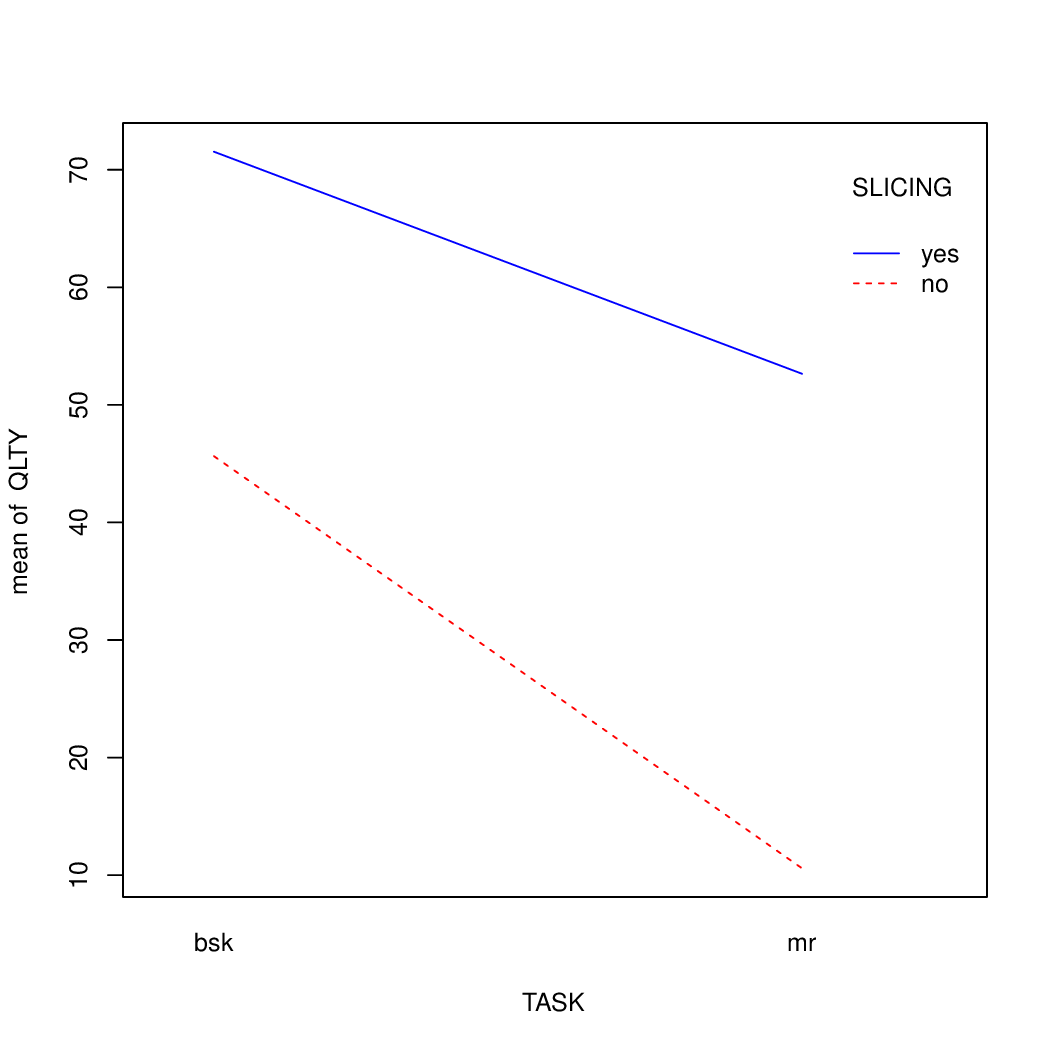}
\caption{Slicing $\times$ Task interaction. (left) ITLD. (right) TDD.}
\label{fig:experiments}
\end{figure}

There is a non-significant $Slicing \times Task$ interaction, displayed in Fig.~\ref{fig:experiments}(left). The dotted line (non-sliced specification) suggests that MR is easier to code than BSK (QLTY scores are higher for MR than BSK). When using a sliced specification, the situation does not change substantially; however, the scores for BSK increase and the scores for MR decrease. This suggests that some tasks benefit to some degree from a sliced specification even in non-TDD situations. This result is coherent with the observations by M\"uller et al.'s \cite{muller2002two}.

% latex table generated in R 3.3.2 by xtable 1.8-2 package
% Thu Aug 23 16:00:49 2018
\begin{table}[ht]
\centering
\caption{Analysis of the effect of the SLICED specification on ITLD programming} 
\label{tab:experiment:itld:analysis}
\begin{tabular}{lrrrrr}
  \hline
 & Df & Sum Sq & Mean Sq & F value & Pr($>$F) \\ 
  \hline
TASK & 1.00 & 1232.35 & 1232.35 & 1.19 & 0.29 \\ 
  SLICING & 1.00 & 15.12 & 15.12 & 0.01 & 0.91 \\ 
  Residuals & 17.00 & 17661.23 & 1038.90 &  &  \\ 
   \hline
\end{tabular}
\end{table}

\subsubsection{TDD phase} Table~\ref{tab:experiment:tdd:descriptives} reports the descriptive statistics for the TDD session. Although the QLTY range is wide ($min: 10.57,\ max: 71.54$), the \textit{Slicing} level \textit{yes} gives higher QLTY scores than the level \textit{no} consistently. Standard deviations are relatively small as compared to the means. The data suggest that the \textit{Slicing} factor has some influence on the QLTY scores.

% latex table generated in R 3.3.2 by xtable 1.8-2 package
% Thu Aug 23 16:00:54 2018
\begin{table}[ht]
\centering
\caption{Descriptive statistics for the TDD session} 
\label{tab:experiment:tdd:descriptives}
\begin{tabular}{rllrrr}
  \hline
 & SLICING & TASK & n & mean & sd \\ 
  \hline
1 & no & bsk & 6.00 & 45.64 & 26.07 \\ 
  2 & yes & bsk & 4.00 & 71.54 & 6.34 \\ 
  3 & no & mr & 5.00 & 10.57 & 10.45 \\ 
  4 & yes & mr & 5.00 & 52.64 & 26.99 \\ 
   \hline
\end{tabular}
\end{table}

Table~\ref{tab:experiment:tdd:analysis} shows the analysis results. H2\textsubscript{0} is rejected. The \textit{Slicing} factor has a statistically significant, very large \cite{cohen2013statistical} positive effect (Cohen's $d = 1.30$, confidence interval $(0.26, 2.33)$). 

In contrast to the ITLD session, there is not a $Slicing \times Task$ interaction, as it can be seen in Fig.~\ref{fig:experiments}(right). Both BSK and MR are easier to code using sliced specifications.

% latex table generated in R 3.3.2 by xtable 1.8-2 package
% Thu Aug 23 16:00:54 2018
\begin{table}[ht]
\centering
\caption{Analysis of the effect of the SLICED specification on TDD programming} 
\label{tab:experiment:tdd:analysis}
\begin{tabular}{lrrrrr}
  \hline
 & Df & Sum Sq & Mean Sq & F value & Pr($>$F) \\ 
  \hline
TASK & 1.00 & 2974.56 & 2974.56 & 7.03 & 0.02 \\ 
  SLICING & 1.00 & 5713.89 & 5713.89 & 13.51 & 0.00 \\ 
  Residuals & 17.00 & 7188.16 & 422.83 &  &  \\ 
   \hline
\end{tabular}
\end{table}

\subsection{Inferences}

When the experimental subjects \textbf{use the sliced specifications during the TDD programming session, the quality of the generated code increases}. As a $Slicing \times Task$ interaction does not exist, the cause of the positive effect should be traced back to the differences between the specifications, i.e., \textbf{the vertical slices}. 

\textbf{The beneficial influence of the vertical slices is restricted to TDD programming}. In the ITLD session, the sliced specification do not have any positive influence and, in fact, the non-sliced specification achieves higher QLTY scores.

Summarizing: The experimental results suggest that \textbf{slicing do have a specific influence on TDD}. 

\subsection{Threats to validity} \label{sec:experiments:threats} 
\begin{figure}
\centering
\includegraphics[width=1.7in]{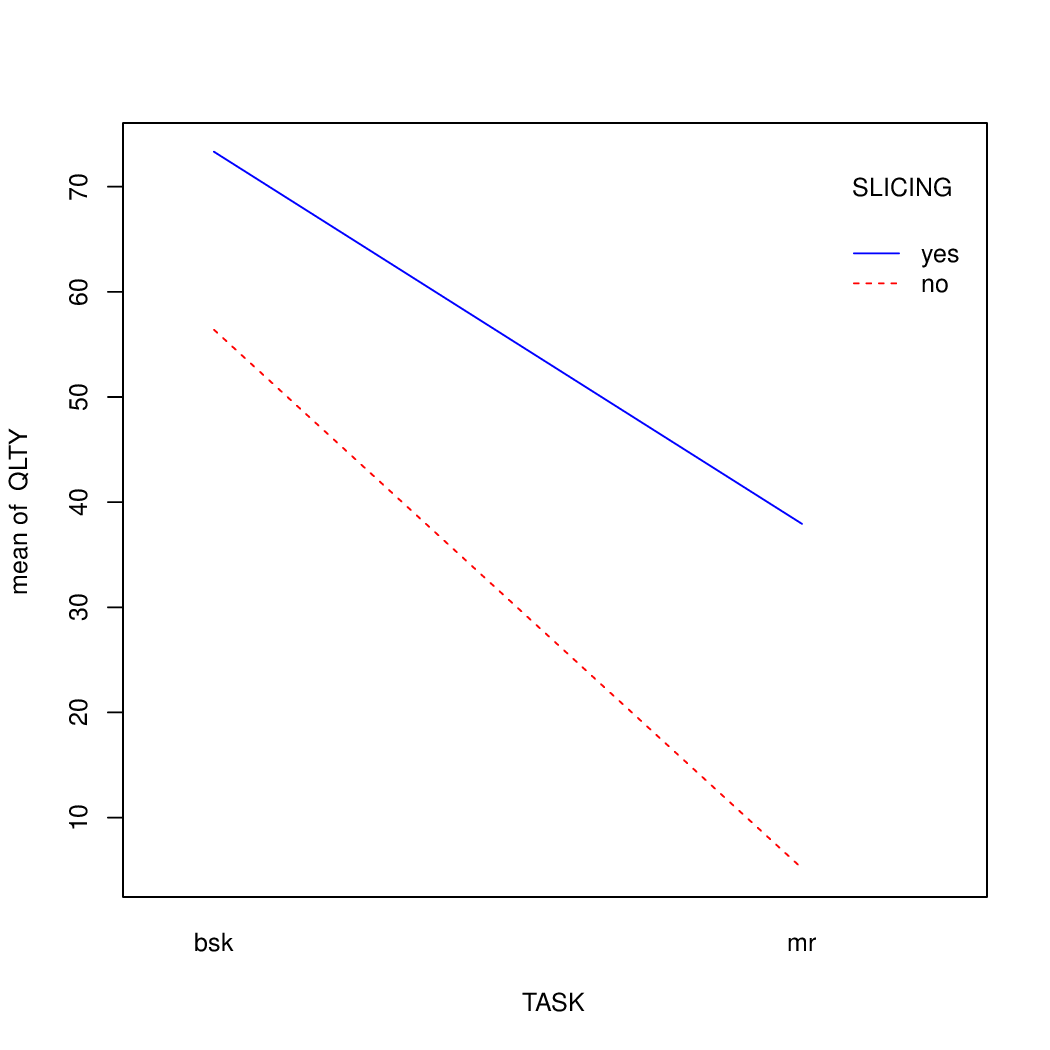}
\includegraphics[width=1.7in]{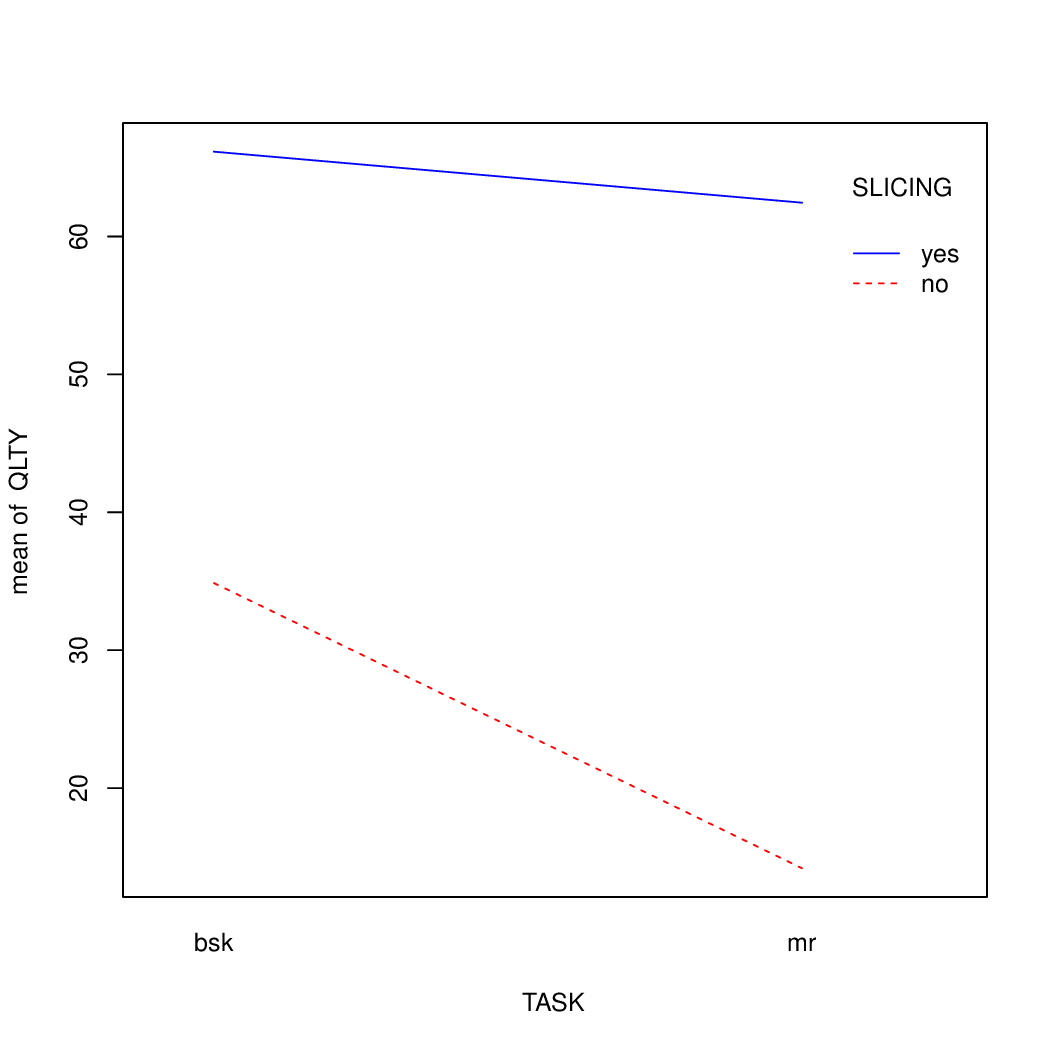}
\caption{Programmers' productivity in TDD. (left) $Slicing \times Task$ interaction considering only subjects that used different types of specifications in the 1\textsuperscript{st} and 2\textsuperscript{nd} sessions. (right) Using the same same type of specification.}
\label{fig:experiments:tdd}
\end{figure}

This experiment is affected by the usual threats to validity that operate in experiments with human subjects. In particular, we wish to emphasize six threats and mitigations strategies:
\begin{enumerate}
\item Statistical power: The sample was too small to achieve a powerful statistical analysis. However, preliminary evidence, e.g., \cite{tosun2017industry} suggested that the effect of the sliced specification was rather large.
\item The ANOVA residuals are non-normal. ANOVA is robust to non-normality \cite{glass1972consequences, lix1996revisited}. However, we have checked using the Wilcoxon test that \textit{Slicing} is:
\begin{enumerate*}[label={\alph*)}]
\item non-significant for ITLD ($W=49,\ p-value=1$),
\item significant for TDD ($W=19,\ p-value=0.02$)
\end{enumerate*}.
\item The experimental subjects were instructed to use a ITLD strategy during the 1\textsuperscript{st} experimental phase. However, it is highly unlikely that they did so. The subjects had almost no \textit{Boost Test} experience, so it was hard for them to follow a disciplined programming approach. 
%In fact, the number of test cases generated in the ITLD programming session was fairly small (\textcolor{red}{X cases in average, as compared to Y cases in the TDD programming session}). 
In our opinion, the subjects solved the ITLD task ''their way''. Far from being a threat to validity, the confidence on the results of this experiment is not affected, or even improves, due to the \textit{natural} way in which the ITLD task was performed.
\item Learning (/maturation): The use of the sliced specification could introduce a bias. The characteristics of this bias, e.g., learning rate, are uncertain. We decided to counter-balance the assignment of sliced specifications to subjects during the 2\textsuperscript{nd} experimental phase.
\item Counter-balancing is another threat to validity in itself. However, we have checked that the beneficial influence of sliced specifications holds both for the subjects that used the different (Fig.~\ref{fig:experiments:tdd}-left) or the same (Fig.~\ref{fig:experiments:tdd}-right) type of specification in the 1\textsuperscript{st} and 2\textsuperscript{nd} sessions.
\item The sliced specification could be designed to increase QLTY purposely. However, notice that the sliced specification of BSK was created by Hakan Erdogmus and used in Erdogmus et al. \cite{erdogmus2005effectiveness}, long time before this research took shape (Ericsson's experiment was conducted in 2015). The specification of MR was created by Oscar Dieste, mimicking BSK specification. MR achieved lower quality scores than BSK.
\end{enumerate}

\section{Conclusions}

TDD is a widely used programming approach. However, we have a limited knowledge about how it works. In this paper, we have made a proposal about the mechanisms underlying TDD. We claim that slicing is at the heart of TDD, namely:
\begin{itemize}
\item Each TDD cycle represents a (nano) vertical slice of some (probably also small) user story.
\item The vertical slice is captured (we could said, ''specified'') using a contract (not explicit, but implicit in the developer's mind).
\item The code created during a TDD cycle is a sliced-based specification, using the contract as slicing pre/post-condition.
\end{itemize}

We have provided evidence that the connections between TDD, contracts and slices are plausible by means of a controlled experiment, conducted at industry with professional developers. 

Nevertheless, there are several testable hypotheses following the previous claims. These hypotheses represent lines of future research: 
\begin{enumerate*}[label={\alph*)}]
\item First and foremost, we need to check if experienced TDD programmers really organize the TDD cycles as contract-based slices. 
\item Secondarily, we need to verify whether slices are combined, or not,  using logical OR, and the impact of the combination strategies in the test cases.
\end{enumerate*}
Conducting this research is challenging, because experienced TDD programmers are in short supply. Crowdsourcing, or hackatons, are options to recruit experimental subjects. We are working on both directions.

Another interesting future research line relates to the correctness character of TDD. We wonder whether programmers could benefit from more powerful programming environments, where in addition to testing, mocking and refactoring, other options like slicing (contracts are taken for granted), or proof assistants, were available. The amount of research in this area is really huge, and we will have to catch up before proceeding.

\section*{Acknowledgment}
We would like to acknowledge Dr. Hakan Erdogmus who contributed to the design of one of the tasks used in the study (BSK) and the corresponding test cases. We also wish to acknowledge Mr. Timo Raty for his participation in the creation of the code templates for C++.

% Notice that it is located before the references - this is not standard
%\appendix

%\input{appendix-div-statement} % the idea was to include this as an appendix

\clearpage

% References
\bibliographystyle{IEEEtran}
\bibliography{paper}

\begin{table*}[]
\centering
\caption{TDD session for the \lstinline{div()} function}
\label{tab:div-listing1}
\begin{adjustbox}{max totalheight=\textheight, keepaspectratio}
\begin{tabular}{lll}
\multicolumn{1}{c}{\textit{Test}} & \multicolumn{1}{c}{\textit{Slice}} & \multicolumn{1}{c}{\textit{Code}} \\ \hline
% 1st row
\begin{minipage}{.38\linewidth}
\begin{lstlisting}
#define BOOST_TEST_MAIN
#include <boost/test/included/unit_test.hpp>

#include "div.hpp"
BOOST_AUTO_TEST_CASE (divide_2_by_2)
{
	const int DIVIDEND = 2;
	const int DIVISOR = 2;
	const int QUOTIENT = 1;
	const int REMAINDER = 0;
	int quotient;
	int remainder;
	
	div(DIVIDEND, DIVISOR, &quotient, &remainder);
	BOOST_CHECK_EQUAL(quotient, QUOTIENT);
	BOOST_CHECK_EQUAL(remainder, REMAINDER);
}
\end{lstlisting}
\end{minipage}
& \multicolumn{1}{l|}{} 
\begin{minipage}{.01\linewidth}
\end{minipage}
\begin{minipage}{.36\linewidth}
\begin{lstlisting}[mathescape=true]
$P:\{x = 2 \land y = 2\}$
void div(int x, int y, 
		 int* q, int* r) {
	int t;
	
	t = x;
	*q = 0;
$\{x = 2 \land y = 2\ \land t = 2 \land q = 0\}$
	if(t >= y) { // unrolled loop
		t = t - y;
		(*q)++;
	}
$\{x = 2 \land y = 2\ \land t = 0 \land q = 1\}$
	¡if(t >= y) {¡
		¡t = t - y;¡
		¡(*q)++;¡
	¡}¡
$\{x = 2 \land y = 2\ \land t = 0 \land q = 1\}$
	¡if(t >= y) {¡
		¡t = t - y;¡
		¡(*q)++;¡
	¡}¡
$\{x = 2 \land y = 2\ \land t = 0 \land q = 1\}$
	*r = t;
$\{x = 2 \land y = 2\ \land t = 0 \land q = 1 \land r = 0\}$
}
$Q:\{0 \leq r < y \land x=y*q+r\}$
\end{lstlisting}
\end{minipage}
& \multicolumn{1}{l|}{}        
\begin{minipage}{.01\linewidth}
\end{minipage}
\begin{minipage}{.24\linewidth}
\begin{lstlisting}
void div(int x, int y, 
		 int* q, int* r) {
	*q = 1;
	*r = 0;
}
\end{lstlisting}
\end{minipage} \\ \hline
% 2nd row
\begin{minipage}{.38\linewidth}
\begin{lstlisting} 
BOOST_AUTO_TEST_CASE (divide_4_by_2)
{
	const int DIVIDEND = 4;
	const int DIVISOR = 2;
	const int QUOTIENT = 2;
	const int REMAINDER = 0;
	int quotient;
	int remainder;
	
	div(DIVIDEND, DIVISOR, &quotient, &remainder);
	BOOST_CHECK_EQUAL(quotient, QUOTIENT);
	BOOST_CHECK_EQUAL(remainder, REMAINDER);
}
\end{lstlisting}
\end{minipage}
& \multicolumn{1}{l|}{} 
\begin{minipage}{.01\linewidth}
\end{minipage}
\begin{minipage}{.36\linewidth}
\begin{lstlisting}[mathescape=true]
$P:\{(x = 2 \lor x = 4) \land y = 2\}$
void div(int x, int y, 
		 int* q, int* r) {
	int t;
	
	t = x;
	*q = 0;
$\{(x = 2 \lor x = 4) \land y = 2\ \land (t = 2 \lor t = 4)  \land q = 0\}$	
	if(t >= y) { // unrolled loop
		t = t - y;
		(*q)++;
	}
$\{(x = 2 \lor x = 4) \land y = 2\ \land (t = 0 \lor t = 2)  \land q = 1\}$
	if(t >= y) {
		t = t - y;
		(*q)++;
	}
$\{(x = 2 \lor x = 4) \land y = 2\ \land t = 0 \land (q = 1 \lor q = 2)\}$
	¡if(t >= y) {¡
		¡t = t - y;¡
		¡(*q)++;¡
	¡}¡
$\{(x = 2 \lor x = 4) \land y = 2\ \land t = 0 \land (q = 1 \lor q = 2)\}$
	*r = t;
$\{(x = 2 \lor x = 4) \land y = 2\ \land t = 0 \land (q = 1 \lor q = 2)$$\land r = 0\}$
}
$Q:\{0 \leq r < y \land x=y*q+r\}$
\end{lstlisting}
\end{minipage}
& \multicolumn{1}{l|}{}        
\begin{minipage}{.01\linewidth}
\end{minipage}
\begin{minipage}{.24\linewidth}
\begin{lstlisting}
void div(int x, int y, 
		 int* q, int* r) {
	int t;

	t = x;
	*q = 1;
	if(t > 0) {
		t = t - y;
		(*q)++;
	}
	*r = 0;
}
\end{lstlisting}
\end{minipage} \\ \hline 
% 3rd row
\begin{minipage}{.38\linewidth}
\begin{lstlisting} 
BOOST_AUTO_TEST_CASE (divide_6_by_2)
{
	const int DIVIDEND = 6;
	const int DIVISOR = 2;
	const int QUOTIENT = 3;
	const int REMAINDER = 0;
	int quotient;
	int remainder;
	
	div(DIVIDEND, DIVISOR, &quotient, &remainder);
	BOOST_CHECK_EQUAL(quotient, QUOTIENT);
	BOOST_CHECK_EQUAL(remainder, REMAINDER);
}
\end{lstlisting}
\end{minipage}
& \multicolumn{1}{l|}{} 
\begin{minipage}{.01\linewidth}
\end{minipage}
\begin{minipage}{.36\linewidth}
\begin{center}
The new contract is:

$P:\{x = 2^n \land 1 \leq n \leq 3 \land y = 2\}$
$Q:\{0 \leq r < y \land x=y*q+r\ \land r = 0\}$

The computation of the slice is similar; the third loop cannot be removed.
\end{center}
\end{minipage}
& \multicolumn{1}{l|}{}        
\begin{minipage}{.01\linewidth}
\end{minipage}
\begin{minipage}{.24\linewidth}
\begin{lstlisting}
void div(int x, int y, 
		 int* q, int* r) {
	int t;

	t = x;
	*q = 0;
	if(t > 0) {
		t = t - y;
		(*q)++;
	}
	if(t > 0) {
		t = t - y;
		(*q)++;
	}
	*r = 0;
}
\end{lstlisting}
\end{minipage} \\ \hline 
% 4rd row
\begin{minipage}{.38\linewidth}
Refactoring, followed by regression test
\end{minipage}
& \multicolumn{1}{l|}{} 
\begin{minipage}{.01\linewidth}
\end{minipage}
\begin{minipage}{.36\linewidth}
\begin{lstlisting}[mathescape=true]
$P:\{x = 2^n \land n > 0 \land y = 2\}$
void div(int x, int y, 
		 int* q, int* r) {
	int t;
$\{x=y^n \land n > 0\}$
	t = x;
$\{x=t \land 0 \leq t \leq x \land t = y^n \land n > 0\}$
	*q = 0;
$\{x=y*q+t) \land 0 \leq t \leq x \land t = y^{n-q} \land n > 0\}$
	while(t >= y) {
		t = t - y;
		(*q)++;
	}
$\{x=y*q+t \land t = 0\}$$\implies$$\{0 \leq t < y \land x=y*q+t \land t = 0\}$
	*r = t;
}
$Q:\{0 \leq r < y \land x=y*q+r \land r = 0\}$
\end{lstlisting}
\end{minipage}
& \multicolumn{1}{l|}{}        
\begin{minipage}{.01\linewidth}
\end{minipage}
\begin{minipage}{.24\linewidth}
\begin{lstlisting}
void div(int x, int y, 
		 int* q, int* r) {
	int t;

	t = x;
	*q = 1;
	while(t > 0) {
		t = t - y;
		(*q)++;
	}
	*r = 0;
}
\end{lstlisting}
\end{minipage} \\ \hline
\end{tabular}
\end{adjustbox}
\end{table*}
%
%
% 2nd part of the table
%
%

\begin{table*}
\centering
\caption{TDD session for the \lstinline{div()} function (cont'd)}
\label{tab:div-listing2}
\begin{adjustbox}{max totalheight=\textheight, keepaspectratio}
\begin{tabular}{lll}
\multicolumn{1}{c}{\textit{Test}} & \multicolumn{1}{c}{\textit{Slice}} & \multicolumn{1}{c}{\textit{Code}} \\ \hline
% 5th row
\begin{minipage}{.38\linewidth}
\begin{lstlisting} 
BOOST_AUTO_TEST_CASE (divide_0_by_2)
{
	const int DIVIDEND = 0;
	const int DIVISOR = 2;
	const int QUOTIENT = 0;
	const int REMAINDER = 0;
	int quotient;
	int remainder;
	
	div(DIVIDEND, DIVISOR, &quotient, &remainder);
	BOOST_CHECK_EQUAL(quotient, QUOTIENT);
	BOOST_CHECK_EQUAL(remainder, REMAINDER);
}
\end{lstlisting}
\end{minipage}
& \multicolumn{1}{l|}{} 
\begin{minipage}{.01\linewidth}
\end{minipage}
\begin{minipage}{.36\linewidth}
\begin{lstlisting}[mathescape=true]
$P:\{x = 2^n \land n \geq 0 \land y = 2\}$
void div(int x, int y, 
		 int* q, int* r) {
	int t;
$\{x=y^n \land n \geq 0\}$
	t = x;
$\{x=t \land 0 \leq t \leq x \land t = y^n \land n \geq 0\}$
	*q = 0;
$\{x=y*q+t) \land 0 \leq t \leq x \land t = y^{n-q} \land n \geq 0\}$
	while(t >= y) {
		t = t - y;
		(*q)++;
	}
$\{x=y*q+t \land t = 0\}$$\implies$$\{0 \leq t < y \land x=y*q+t \land t = 0\}$
	*r = t;
}
$Q:\{0 \leq r < y \land x=y*q+r \land r = 0\}$
\end{lstlisting}
\end{minipage}
& \multicolumn{1}{l|}{}        
\begin{minipage}{.01\linewidth}
\end{minipage}
\begin{minipage}{.24\linewidth}
\begin{lstlisting}
void div(int x, int y, 
		 int* q, int* r) {
	int t;

	t = x;
	*q = 0;
	while(t > 0) {
		t = t - y;
		(*q)++;
	}
	*r = 0;
}
\end{lstlisting}
\end{minipage} \\ \hline 
% 6th row
\begin{minipage}{.38\linewidth}
\begin{lstlisting} 
BOOST_AUTO_TEST_CASE (divide_4_by_1)
{
	const int DIVIDEND = 4;
	const int DIVISOR = 1;
	const int QUOTIENT = 4;
	const int REMAINDER = 0;
	int quotient;
	int remainder;
	
	div(DIVIDEND, DIVISOR, &quotient, &remainder);
	BOOST_CHECK_EQUAL(quotient, QUOTIENT);
	BOOST_CHECK_EQUAL(remainder, REMAINDER);
}
\end{lstlisting}
\end{minipage}
& \multicolumn{1}{l|}{} 
\begin{minipage}{.01\linewidth}
\end{minipage}
\begin{minipage}{.36\linewidth}
\begin{center}
The new contract is:

$P:\{(x = y * q\}$
$Q:\{0 \leq r < y \land x=y*q+r \land r = 0\}$

The computation of the slice is similar.
\end{center}
\end{minipage}
& \multicolumn{1}{l|}{}        
\begin{minipage}{.01\linewidth}
\end{minipage}
\begin{minipage}{.24\linewidth}
\begin{lstlisting}
void div(int x, int y, 
		 int* q, int* r) {
	int t;

	t = x;
	*q = 0;
	while(t > 0) {
		t = t - y;
		(*q)++;
	}
	*r = 0;
}
\end{lstlisting}
\end{minipage} \\ \hline 
% 7th row
\begin{minipage}{.38\linewidth}
\begin{lstlisting} 
// triangulation test
BOOST_AUTO_TEST_CASE (divide_9_by_3)
{
	const int DIVIDEND = 9;
	const int DIVISOR = 3;
	const int QUOTIENT = 3;
	const int REMAINDER = 0;
	int quotient;
	int remainder;
	
	div(DIVIDEND, DIVISOR, &quotient, &remainder);
	BOOST_CHECK_EQUAL(quotient, QUOTIENT);
	BOOST_CHECK_EQUAL(remainder, REMAINDER);
}
\end{lstlisting}
\end{minipage}
& \multicolumn{1}{l|}{} 
\begin{minipage}{.01\linewidth}
\end{minipage}
\begin{minipage}{.36\linewidth}
\center{Same slice}
\end{minipage}
& \multicolumn{1}{l|}{}        
\begin{minipage}{.01\linewidth}
\end{minipage}
\begin{minipage}{.24\linewidth}
\begin{lstlisting}
void div(int x, int y, 
		 int* q, int* r) {
	int t;

	t = x;
	*q = 0;
	while(t > 0) {
		t = t - y;
		(*q)++;
	}
	*r = 0;
}
\end{lstlisting}
\end{minipage} \\ \hline 
% 8th row
\begin{minipage}{.38\linewidth}
\begin{lstlisting} 
BOOST_AUTO_TEST_CASE (divide_7_by_2)
{
	const int DIVIDEND = 7;
	const int DIVISOR = 2;
	const int QUOTIENT = 3;
	const int REMAINDER = 1;
	int quotient;
	int remainder;
	
	div(DIVIDEND, DIVISOR, &quotient, &remainder);
	BOOST_CHECK_EQUAL(quotient, QUOTIENT);
	BOOST_CHECK_EQUAL(remainder, REMAINDER);
}
\end{lstlisting}
\end{minipage}
& \multicolumn{1}{l|}{} 
\begin{minipage}{.01\linewidth}
\end{minipage}
\begin{minipage}{.36\linewidth}
\begin{lstlisting}[mathescape=true]
$P:\{x=y*q+r\}$
void div(int x, int y, 
		 int* q, int* r) {
	int t;
$\{\top\}$
	t = x;
	*q = 0;
$\{x=y*q+t) \land 0 \leq t\}$
	while(t >= y) {
		t = t - y;
		(*q)++;
	}
$\{x=y*q+t \land 0 \leq t \land t < y\}$$\implies$$\{0 \leq t < y \land x=y*q+t\}$
	*r = t;
}
$Q:\{0 \leq r < y \land x=y*q+r\}$
\end{lstlisting}
\end{minipage}
& \multicolumn{1}{l|}{}        
\begin{minipage}{.01\linewidth}
\end{minipage}
\begin{minipage}{.24\linewidth}
\begin{lstlisting}
void div(int x, int y, 
		 int* q, int* r) {
	int t;
	
	t = x;
	*q = 0;
	while(t >= y) {
		t = t - y;
		(*q)++;
	}
	*r = t;
}
\end{lstlisting}
\end{minipage} \\ \hline 
% 9th row
\begin{minipage}{.38\linewidth}
\begin{lstlisting} 
// triangulation test
BOOST_AUTO_TEST_CASE (divide_2_by_9)
{
	const int DIVIDEND = 2;
	const int DIVISOR = 9;
	const int QUOTIENT = 0;
	const int REMAINDER = 2;
	int quotient;
	int remainder;
	
	div(DIVIDEND, DIVISOR, &quotient, &remainder);
	BOOST_CHECK_EQUAL(quotient, QUOTIENT);
	BOOST_CHECK_EQUAL(remainder, REMAINDER);
}
\end{lstlisting}
\end{minipage}
& \multicolumn{1}{l|}{} 
\begin{minipage}{.01\linewidth}
\end{minipage}
\begin{minipage}{.36\linewidth}
\center{Same slice}
\end{minipage}
& \multicolumn{1}{l|}{}        
\begin{minipage}{.01\linewidth}
\end{minipage}
\begin{minipage}{.24\linewidth}
\begin{lstlisting}
void div(int x, int y, 
		 int* q, int* r) {
	int t;
	
	t = x;
	*q = 0;
	while(t >= y) {
		t = t - y;
		(*q)++;
	}
	*r = t;
}
\end{lstlisting}
\end{minipage} \\ \hline
\end{tabular}
\end{adjustbox}
\end{table*}

\end{document}